\documentclass[11pt,twoside]{article}

\usepackage[a4paper,hmarginratio=1:1,vmarginratio=2:3,totalwidth=15.6cm,
totalheight=20.5cm]{geometry}

\usepackage{amssymb,cite,axodraw}
\usepackage{epsfig}
\usepackage{latexsym}

\newcommand{\ra}{\rightarrow}

\newcommand{\cJ}{{\cal J}}
\newcommand{\cO}{{\cal O}}

\newcommand{\cL}{{\cal L}}

\newcommand{\cV}{{\cal V}}
\newcommand{\cT}{{\cal T}}
\newcommand{\Li}{\mbox{Li}}

\newcommand{\bR}{{\bf R}}
\newcommand{\bZ}{{\bf Z}}

\newcounter{oldcounter}
\addtocounter{equation}{1}
\newcommand{\be}{\begin{equation}}
\newcommand{\ee}{\end{equation}}
\newcommand{\bea}{\begin{eqnarray}}
\newcommand{\eea}{\end{eqnarray}}

\begin{document}

\thispagestyle{empty}
\begin{flushright}
DAMTP-2005-71\\
{DESY-05-144}
\end{flushright}
\vskip 1. cm
\begin{center}
{\Large
{\bf Higher Derivative Operators from }}\\
\vspace{0.3cm}
{\Large{\bf Scherk-Schwarz Supersymmetry Breaking on $\cT^2/Z_2$. \\}}
\vspace{1.cm}
{\bf D.~M. Ghilencea $^{a,}$\footnote{E-mail:
    D.M.Ghilencea@damtp.cam.ac.uk}}
 \,\,and\,\, {\bf Hyun Min Lee $^{b, }$\footnote{E-mail: minlee@mail.desy.de}}\\
\vspace{0.8cm}
{\it $^a $D.A.M.T.P., Centre for Mathematical Sciences,
University of Cambridge, \\
Wilberforce Road, Cambridge CB3 OWA, United Kingdom}\\
\vspace{0.4cm}
{\it $^b $Deutsches Elektronen-Synchrotron DESY, \\
D-22603 Hamburg, Germany}\\
\end{center}
\vspace{0.2cm}
\begin{abstract}
\noindent
In  orbifold compactifications on 
$\cT^2/Z_2$ with Scherk-Schwarz  supersymmetry breaking,
it is shown that (brane-localised) superpotential interactions 
and (bulk) gauge interactions generate at one-loop higher 
derivative counterterms to the mass of the brane (or zero-mode of the 
bulk) scalar field.  
These brane-localised operators are generated by integrating out 
the bulk modes of the initial theory which, 
although supersymmetric, is nevertheless non-renormalisable.
It is argued that such operators, of non-perturbative origin and
 not protected by non-renormalisation  theorems, are generic in orbifold
compactifications and play a crucial  role in the UV behaviour of 
the two-point Green function of  the scalar field self-energy.
Their presence in the action with   unknown coefficients
prevents one from making predictions about physics at (momentum) 
scales close to/above the compactification scale(s).
Our results extend to the case of two dimensional orbifolds, previous 
findings for  $S^1/Z_2$ and $S^1/(Z_2\times Z_2')$ compactifications
where brane-localised higher derivative operators are also dynamically  
generated at loop level, regardless of the details of the 
supersymmetry breaking mechanism. We stress the importance of these
operators for the hierarchy and the cosmological constant
problems in compactified theories.
\end{abstract}

\setcounter{page}{0}

\newpage

\section{Introduction}

In recent years models for physics beyond the Standard Model 
 with additional space dimensions \footnote{For early works on the possibility
of TeV-scale extra dimension, see \cite{antoniadis}.} 
have drawn much attention from the
physics community. In particular  the supersymmetric versions of such 
models  have been  especially popular due to the possibility of 
understanding the mechanism of transmitting the  supersymmetry
breaking in a geometric way  \cite{susyb,equiv,nomura,kl} and of  
generating radiative  electroweak symmetry  breaking  due to extra 
dimensions \cite{Delgado:2001si}.

If supersymmetry is assumed to be broken in a hidden
sector\footnote{For early works on this topic  see \cite{Derendinger:1985kk}.}
that is geometrically separated in extra dimensions or that
does not couple directly to the visible sector, then
there are  no  soft mass terms at the tree level, in the visible sector
\footnote{In the six-dimensional case
the sequestering mechanism does not seem to apply \cite{fll}. 
One can have a nonzero gravity mediation contribution to the soft mass
at the tree level,  in some moduli stabilisation mechanism \cite{fll}.}. 
If so,  non-zero soft mass parameters  are  nevertheless 
generated by the loop corrections,  which  provide in such case  the 
leading contribution.
Further, it has been shown \cite{equiv,hmlee} 
that the codimension-one localised sources of supersymmetry breaking on
orbifolds are equivalent to the Scherk-Schwarz  breaking due to
non-trivial boundary conditions.
Then,  in the presence of either local or Scherk-Schwarz 
breaking of supersymmetry,  one can compute the loop-corrections to the 
masses of the scalar fields in the visible sector.
As an application, in some models with one extra dimension 
compactified on $S^1/Z_2$ 
or $S^1/(Z_2\times Z'_2)$ orbifolds,  the one-loop correction to the
mass of a scalar (Higgs) field may be UV cutoff independent and
 have a negative sign to trigger
(electroweak)  symmetry breaking \cite{Delgado:2001si,higgsmass}.  
However, the situation is in general more complicated \cite{Ghilencea:2001bw,
Ghilencea:2004sq,Ghilencea:2005hm}.

It has been shown recently that in 
$S^1/Z_2$ and $S^1/(Z_2\!\times\! Z_2')$ orbifolds, higher derivative
operators are dynamically generated, already  at the one-loop level,
 as  counterterms to the  mass of the scalar field 
(usually identified with the Higgs field)  
\cite{Ghilencea:2004sq,Ghilencea:2005hm}.
This happens whether the scalar field is a brane field  or a zero-mode 
of a bulk field, and it was shown to be present regardless
of the way  supersymmetry was broken\footnote{There is an
exception in the case of  F-term breaking, where a footprint
of this breaking mechanism is still manifest as a coefficient in front
of the higher derivative operators,  but these are nevertheless
 generated \cite{Ghilencea:2005hm}.}:
local (F-term) breaking, (non-local) discrete and continuous
 Scherk-Schwarz mechanism or additional orbifolding ($Z_2'$).
In fact the presence of such operators has little or no dependence
on the particular choice of the supersymmetry
breaking mechanism, and
is actually due to the number of bulk fields involved in the
interaction at the loop level.
Therefore,  the presence of higher derivative operators is generic 
in theories with extra dimensions. Their implications
for studies of (electroweak) symmetry breaking, for the
 UV behaviour of these theories, and for the hierarchy problem in
particular,  must then  be carefully investigated. Our
previous findings in \cite{Ghilencea:2004sq,Ghilencea:2005hm}   suggested
that the initial supersymmetry of the higher dimensional theory cannot
prevent the emergence of such operators at the radiative level.
This  also raises intriguing questions on the role of initial supersymmetry
in ensuring a mild UV ``running'' of the loop-corrected mass of the
scalar (Higgs) field at scales of order  $1/R^2$ where such operators 
become relevant.

The presence of higher derivative counterterms to the masses
of scalar fields is closely related to the number of
Kaluza-Klein towers to sum over in the dimensional reduction of the
action  at the loop level. This number  is increased in the
case of loop corrections from (brane-localised) interactions which do
not respect momentum conservation in the extra dimensions. For a fixed
order in perturbation theory this makes  more likely the generation of
such operators from (brane localised) Yukawa interactions  than   
from (bulk) gauge interactions. For similar reasons
 higher derivative operators also emerge  from  compactification
as counterterms to the gauge kinetic terms \cite{Oliver:2003cy}.

These findings are ultimately related to the non-renormalisable
character of the higher dimensional theories which becomes manifest 
above the compactification scale(s).
Although supersymmetric, such (effective) theories still have some 
of the shortcomings of the non-renormalisable theories, such as 
unknown UV behaviour, controlled by the coefficients of  higher dimension 
operators. For the  case of higher (dimension) derivative operators
more complications arise due to the introduction  of extra
degrees of freedom (ghost fields), possible unitarity violation,
non-locality effects, which made such theories less popular in the
past \cite{Pais}.

This paper aims to  extend the validity of the above findings 
about higher derivative counterterms to the scalar field mass in 
one-dimensional  orbifolds \cite{Ghilencea:2004sq,Ghilencea:2005hm},  
to the case of two dimensional orbifolds. 
For a ${\cal T}^2/Z_2$ compactification
we shall  consider the effects of the  localised Yukawa interaction 
at the orbifold fixed point and  also of the (bulk) gauge interaction
on the one-loop correction to the mass of a scalar field. In 
many applications this field plays the role of the
Higgs field. The conclusion is that both Yukawa and gauge interactions 
generate - at the one-loop level - higher derivative counterterms to
the mass of the scalar field. This indicates that, although little
investigated in the past, higher derivative  counterterms are, rather
interestingly, a generic presence in  orbifold compactifications, 
at the quantum level, and are not protected by non-renormalisation
theorems.

The paper is organised as follows.
In the next section  we outline the setup of the model that we are considering.
Then we derive the Kaluza-Klein spectrum of the  bulk fields (vector
multiplets and hypermultiplets) under the boundary conditions of the
orbifold and of the Scherk-Schwarz supersymmetry breaking  on 
${\cal T}^2/Z_2$. 
In the presence of this  breaking Section~\ref{section4} provides 
the details of the calculation of the one-loop correction to the mass
of the scalar field induced  by  the localised Yukawa and (bulk) 
gauge interactions. The conclusions are presented in Section~\ref{section5}.
The Appendix provides the results of evaluating  series of 
integrals  generic in orbifold compactifications,
 which  may be useful in other applications as well.

\section{The model setup.}\label{section2}

We consider a two dimensional compactification on the $\cT^2/Z_2$ orbifold.
The two-torus  is parametrised 
by $z, \bar z$ with  $z=(y_1+iy_2)/2$ and $y_1\in(-\pi R_1,\pi R_1]$,
 $y_2\in(-\pi R_2,\pi R_2]$, and  is invariant under
\bea
z\rightarrow z+(m+nU)\,\pi R_1
\eea
where  $m,n$ are integers and
$U=({R_2}/{R_1}) \,e^{i\theta}\equiv U_1+iU_2$ is the complex
structure of $\cT^2$.
The geometric action of  parity is $Z_2:\,z\rightarrow -z$. 
Therefore there  appear four fixed points which are:
$z=0$, $\pi R_1/2$, $\pi R_1U/2$ and $\pi R_1(1+U)/2$.  

On the orbifold $\cT^2/Z_2$ one can consider vector multiplets 
and hypermultiplets. A vector multiplet is described in a 4D language as
made of a vector superfield  $V(\lambda^1,A_\mu,D^3-F_{56})$ 
and an adjoint chiral
superfield $\Sigma((A_6+i A_5)/\sqrt{2}, \lambda^2,D^1+iD^2)$, where
$\lambda_{1,2}$ are Weyl fermions, $A_\mu, A_5, A_6$ the bulk gauge fields
and $D^i(i=1,2,3)$ the auxiliary fields.
The hypermultiplet contains two
 chiral superfields $\Phi(\phi,\psi,F_\Phi)$ and 
$\Phi^c(\phi^c, \psi^c,F_{\Phi^c})$
with opposite SM quantum numbers, and where $\phi, \phi^c$ are complex
scalars; $\psi,\psi^c$ are Weyl fermions and $F_\Phi,F_{\Phi^c}$ are 
the auxiliary fields. 
We consider now the following parity assignments
\begin{eqnarray}\label{orbifold}
\Phi(x, -z)=\Phi(x,z),\quad  \qquad &&  V(x,-z)=V(x,z),
\nonumber\\
\qquad\Phi^c(x, -z)=-\Phi^c(x,z), \qquad && \Sigma(x,-z)=-\Sigma(x,z),
\end{eqnarray}
where $\Phi$ is a bulk field. Unlike the case of the gauge
 multiplet which is always a bulk field, 
in an orbifold compactification  
not all SM matter fields  $Q, U, D, L, E$ or Higgs fields
 are necessarily bulk fields. 
Thus $\Phi$ may stand only for a subset of these
 fields and this subset will be detailed shortly. 
As a result of eq.(\ref{orbifold}) the original 6D N=1 supersymmetry is broken
and  the fixed points of the orbifold have a remaining
4D N=1 supersymmetry.

Let us consider that the model has a gauge symmetry G.
The action for a hypermultiplet $\Phi$ belonging to a representation
of the gauge group $G$ is \cite{Arkani-Hamed:2001tb}
\bea
{\cal L}_{\rm hyper}\! &=&\!\int\!  dy_1\, dy_2
\bigg\{\int \!\!d^4\! \theta
\,\, \Big[{\overline \Phi}\,e^{2V}\,\Phi
+{\overline \Phi}^c\, e^{-2V}\,\Phi^c\Big] 
+\Big[\int \!\!d^2\! \theta
\,\, \Phi^c\,(-\partial+\sqrt{2}\Sigma)\,\Phi+h.c.\Big]\bigg\}\quad
\eea
with $V=V^a\, T^a_R$ and $\Sigma=\Sigma^a \,T^a_R$ with $T^a_R$ the
group generators. Since not all fields of the model are bulk fields, 
we need the action for a brane chiral multiplet $\Psi$ charged
under the  group $G$, which is 
\bea\label{gauge_int}
{\cal L}_{\rm chiral}=\int\!  dy_1\, dy_2 \; \delta(y_{1,2})
\int \!\!d^4\! \theta \,\,{\overline \Psi}\,e^{2V}\,\Psi
\eea
with $\delta(y_{1,2})\equiv\delta(y_1)\,\delta(y_2)$ with $\delta(y_i)$
the one-dimensional  Dirac delta
function.

A generic presence  for realistic model
building in such compactification
is  a superpotential interaction, which
can only be localised
\begin{eqnarray}
  \cL_Y = \int\!  dy_1\, dy_2 \; \delta(y_{1,2})
  \left\{\int \!\!d^2\! \theta
 \,\, \Big[\lambda_t\, Q \,U \,H_u
      + \lambda_b \, Q \, D\,  H_d + \cdots \Big]
  + {\rm h.c.} \right\}. \label{interaction}\\[-13pt] \nonumber
\end{eqnarray}
The 6D coupling
$\lambda_{t}=f_{6, t}/M_*^n=\cV^{n/2} f_{4, t}$ where $f_{6,t}$
 ($f_{4, t}$) is  the dimensionless 6D  (4D),
$M_*$ is the UV cutoff of the theory and 
$\cV=(2\pi)^2 R_1 R_2 \sin\theta$ is the area of the underlying
two-torus. Dimensional analysis  gives
that $n=1$ if there are two brane fields  and 
one bulk  field in (\ref{interaction}). 
One has  $n=2$ if there is one brane
field in this equation, with 
the other two as bulk fields. 

In the following we shall consider the one-loop effects of this
 superpotential interaction on the mass of the scalar component
 $\phi_{H_{u,d}}$ of $H_{u,d}$. One-loop corrections
from  gauge interactions will also be computed.
For simplicity,  we assume $H_{u,d}$ are brane fields. 
(This is no special restriction: they can also be bulk fields, in which 
case they respect condition (\ref{orbifold}) for hypermultiplets. 
In that case the correction to the mass of the scalar field  will
refer to the zero-mode of $\phi_{H_{u,d}}$). Before proceeding with 
the calculation, one must address, for realistic model building,  
the breaking of the  remaining 4D  N=1 supersymmetry. This is
considered below,  with its implications on the spectrum
 of the bulk fields of the model.

\section{Scherk-Schwarz breaking of supersymmetry on $\cT^2/Z_2$.}\label{section3}

We shall use the continuous Scherk-Schwarz mechanism \cite{ss}
on $\cT^2/Z_2$ orbifold  with complex structure $U$ for the underlying
two-torus, in  order to break the remaining  supersymmetry of the fixed points.
One can also consider other methods of breaking
such as the discrete version of the Scherk-Schwarz
mechanism\footnote{See for example details in  \cite{hmlee}.}, but  we
expect to obtain  similar conclusions.

On the orbifold $\cT^2/Z_2$ 
the orbifold boundary conditions and the Scherk-Schwarz 
twists of the bulk gaugino $\lambda\equiv(\lambda^1,\lambda^2)^T$ 
are as follows,
\bea
Z_2&:&~~\lambda(x,-z)=\sigma_3\, \lambda(x,z)
\equiv P\lambda(x,z), \label{orb}\\[2pt]
T_1&:&~~\lambda(x,z+\pi R_1)=e^{-2\,i \pi\,\omega_1\,\sigma_2}
\lambda(x,z)\equiv T_1\,\lambda(x,z), \label{tw1}\\[2pt]
T_2&:&~~\lambda(x,z+\pi R_1 U)=e^{-2\,i \pi\,\omega_2\,\sigma_2} 
\lambda(x,z)\equiv T_2\,\lambda(x,z) \label{tw2}
\eea
where $\omega_1,\omega_2$ are real (arbitrary) parameters.
Here we note that the consistency conditions $T_i\,P\,T_i=P$, $(i=1,2)$ 
and $T_1\,T_2=T_2\,T_1$ are satisfied.
The study of these boundary conditions on the action for  the gauginos
is easier if  we introduce the untwisted fields $\chi$ defined by
\be\label{redef}
\lambda(x,z)=e^{-i({\bar\alpha}\,z+\alpha\,{\bar z})\sigma_2}\,
\chi(x,z), \qquad {\rm with}\qquad
\alpha=\frac{1}{iR_1U_2}(\omega_1U-\omega_2).
\ee
One can show that $\chi$ satisfies the same orbifold boundary
condition as in eq.~(\ref{orb}), but unlike $\lambda$, $\chi$ is
periodic on the torus. With this re-definition of the fields, 
we can write the gaugino kinetic term   in terms of the
untwisted fields $\chi$ 
\bea\label{gaugeaction}
{\cal L}=\sum_{j=1,2}\Big( 
i\,\chi^j\sigma^\mu\partial_\mu{\bar\chi}^j
+i{\bar\chi}^j {\bar\sigma}^\mu\partial_\mu\chi^j
\Big)
+\Big[-\chi^1\partial_z\chi^2
+\chi^2\partial_z\chi^1+c.c.\Big]
+{\cal L}_m \label{fnl}
\eea
where  
${\cal L}_m$ corresponds to the bulk mass terms given by
\bea
{\cal L}_m=- \Big[ {\bar\alpha}\,
(\chi^1\chi^1+\chi^2\chi^2)+c.c.\Big].\label{twistedmass}
\eea
From the action given in eq.(\ref{fnl}) we derive 
the equations of motion for gauginos 
\bea
i\sigma^\mu\partial_\mu{\bar\chi}^2+\partial_z\chi^1
-{\bar\alpha}\,\chi^2&=&0, \nonumber\\
i{\bar\sigma}^\mu\partial_\mu\chi^1-{\bar \partial_z}{\bar\chi}^2
-\alpha\,{\bar\chi}^1&=&0.
\eea
Solving the above equations gives the solution for the untwisted 
gaugino as 

\be
\left(\begin{array}{l}\chi^1 \\ \chi^2\end{array}\right)(x,z)
=\frac{1}{\sqrt{\cV}}
\sum_{n_1,n_2\in {\bf Z}} 
\left(\begin{array}{l} \cos({\bar c}_{n_1,n_2}z+c_{n_1,n_2}{\bar z}) 
\\ \sin({\bar c}_{n_1,n_2}z+c_{n_1,n_2}{\bar z}) \end{array}\right)
\eta^{(n_1,n_2)}(x)
\ee
where $\cV$  
is the area of  underlying $\cT^2$ and
\bea
c_{n_1,n_2}&=&\frac{1}{iR_1U_2}\,(n_1U-n_2)
\eea
and
$i\,\sigma^\mu\,\partial_\mu\,{\bar\eta}^{(n_1,n_2)}(x)
=M_{n_1,n_2}\eta^{(n_1,n_2)}(x)$. The mass spectrum is given by

\be
{\overline M}_{n_1,n_2}
=\frac{1}{i\,R_1U_2}\,\Big[(n_1+\omega_1)U-(n_2+\omega_2)\Big],
\label{ssmass}
\ee
which, if $U=i\frac{R_2}{R_1}$, simplifies into
\be\label{fmass}
{\overline
  M}_{n_1,n_2}=\frac{\omega_1+n_1}{R_1}+i\bigg(\frac{\omega_2+n_2}{R_2}
\bigg).
\ee
Finally, using relation (\ref{redef}), 
we find the solution for the twisted gaugino $\lambda$ 

\bea
\left(\begin{array}{l}\lambda^1 \\ \lambda^2\end{array}\right)(x,z)
=\frac{1}{\sqrt{\cV}}
\sum_{n_1,n_2\in {\bf Z}}
\left(\begin{array}{l} \cos[({\bar c}_{n_1,n_2}+{\alpha})z
+(c_{n_1,n_2}+\alpha){\bar z}]
\\ \sin[({\bar c}_{n_1,n_2}+{\alpha})z
+(c_{n_1,n_2}+\alpha){\bar z}] 
\end{array}\right)
\eta^{(n_1,n_2)}(x).\label{twistedg}\\
\nonumber
\eea
Eqs.(\ref{ssmass}), (\ref{fmass}) and (\ref{twistedg}) will be used
shortly  in  Section~\ref{gauge-correction} to compute the gauge
corrections to a brane scalar field.

A similar mechanism can be considered for the scalars $(\phi,\phi^c)$ 
belonging to the hypermultiplet $\Phi$. 
Unlike their fermionic partners, they can
acquire (under translation along $y_{1,2}$)
a continuous Scherk-Schwarz phase due to the $SU(2)_R$ symmetry. 
Thus $(\phi,\phi^c)$  respect
conditions similar to (\ref{tw1}), (\ref{tw2}).
The Scherk-Schwarz  phase ``lifts'' the mass of their Kaluza-Klein modes, 
and in particular of their zero-modes which (unlike their fermionic
partners) become massive, to break the remaining 4D N=1 supersymmetry.
Following closely the same steps as for the gaugino fields, 
the mode expansion of scalars is  obtained

\bea
\left(\begin{array}{l}\phi \\[4pt]
 \phi^{c\dagger}\end{array}\right)(x,z)
=\frac{1}{\sqrt{\cV}}
\sum_{n_1,n_2\in {\bf Z}}
\left(\begin{array}{l} \cos[({\bar c}_{n_1,n_2}+{\alpha})z
+(c_{n_1,n_2}+\alpha){\bar z}]
\\[4pt]
\sin[({\bar c}_{n_1,n_2}+{\alpha})z
+(c_{n_1,n_2}+\alpha){\bar z}]
\end{array}\right) \phi_{n_1,n_2}(x)\label{twistsq}\\
\nonumber
\eea
where $(\square-|m_{\phi, n_1,n_2}|^2)\phi_{n_1,n_2}(x)=0$
and 
\begin{eqnarray}\label{massphi}
|m_{\phi, n_1,n_2}|^2=\frac{(2\pi)^2}{\cV \,U_2}\,
|(n_2+\omega_2)-U(n_1+\omega_1)|^2
\end{eqnarray}
This equation will be used in Section~\ref{yukawa-correction} when
computing one loop corrections which involve
the mass of the Kaluza-Klein modes of the bulk scalar fields (squarks).

\section{Higher derivative counterterms on the $\cT^2/Z_2$ orbifold.}
\label{section4}

\subsection{One-loop mass correction from Yukawa interaction.}
 \label{yukawa-correction}

In this section we compute the one-loop correction induced by
interaction (\ref{interaction})  to the  two point Green function of
the self-energy 
of the scalar field $\phi_{H_u}$ (hereafter denoted simply $\phi_H$)
 of $H_u$, which  is considered a brane field. 
The calculation is very similar  if this field is a bulk field
instead, and then $\phi_H$ and its mass correction will refer to the zero 
mode component.
 We restrict the calculation to the first 
interaction in (\ref{interaction}), with a similar approach for the
down-type interaction.
We also consider that in eq.(\ref{interaction}) 
 the quark SU(2) doublets $Q$ are bulk fields while 
the quark singlets $U$ are brane fields. 
This apparent restriction is made to
simplify the one-loop calculation  we perform, to avoid the
proliferation of a large number of associated Kaluza-Klein
sums.  It is obvious that the effects we find and  
which  are ultimately due to the presence of 
 two  Kaluza-Klein summations, will also apply when the $U$ 
fields are also bulk fields \footnote{Although it is possible to
 allow in the bulk  only the $Q$ 
(but not singlets) of one generation, both types of quark (doublet/singlet) 
 of at least another generation must also live in the bulk
for anomaly cancellation \cite{leeanomaly}.}(when further Kaluza-Klein sums are
present). Therefore the restriction above does not affect the
generality of our results.
Finally, interactions of type bulk-boundary-boundary fields that are
considered here are rather  standard in string theory \cite{Hamidi:1986vh}, 
\cite{Antoniadis:1993jp}.

The split multiplet  for the quarks is compatible with orbifold
GUT scenarios.  For example, 
in the 6D version of $SU(5)$ case \cite{kawamura}, 
only the quark doublet comes
from a bulk ${\bf 10}$ as a massless mode\footnote{This possibility 
is considered for the $2^{nd}$
generation quark doublet to explain the $s$-$\mu$ puzzle 
in 5d case \cite{kkl}. }. 
On the other hand, in $SO(10)$ case \cite{abc}, one has massless $Q$ 
and $L$ in $(4,2,1)$ under the Pati-Salam group from two bulk ${\bf 16}$'s.
In this case, there would appear the localised anomalies with split
multiplets which should be cancelled by a localised Green-Schwarz
mechanism \cite{gs}.

In the on-shell formulation, the interaction in eq.~(\ref{interaction})
for the Higgs field $\phi_H$ becomes

\begin{eqnarray}\label{action4d}
  -\cL_Y &=&
    \sum_{k,l\in {\bf Z}}
     \, \,\Bigl[ f_{4, t}\, m_{\phi_Q^c,k,l} \,
    \eta^{F_Q}_{k,l}
    \phi^{c\,\dagger}_{Q,k,l} \,\phi_{U}\, \phi_H
 + {\rm h.c.} \Bigr]
+\sum_{k,l\in {\bf Z}}\,f_{4, t}^2\,(\eta^{F_Q}_{k,l})^2
    \phi^{\dagger}_{U} \phi_{U} \phi^{\dagger}_H \phi_H
\nonumber\\[8pt]
&+&\!\!\!\!\sum_{k,l,m,n\in {\bf Z}}
     \,f_{4, t}^2\,
    \eta^{\phi_Q}_{k,l} \eta^{\phi_Q}_{m,n}
    \phi^{\dagger}_{Q,k,l} \phi_{Q,m,n} \phi^{\dagger}_H \phi_H 
+\!\sum_{k,l\in {\bf Z}}
  \,\Bigl[ f_{4, t}\, \eta^{\psi_Q}_{k,l}
    \psi_{Q,k,l} \psi_{U} \phi_H
  + {\rm h.c.} \Bigr]. \\
\nonumber
\end{eqnarray}
In the above equation 
$\eta^{F_Q}_{m,n}=\eta^{\phi_Q}_{m,n}=\eta^{\psi_Q}_{m,n}=1$.
For a Scherk-Schwarz breaking of supersymmetry,
the spectrum of the squark doublet $\phi_{Q,m,n}$,  $\phi^c_{Q,m,n}$, 
 is  that found in eq.(\ref{massphi}) 
\bea
|m_{\phi_Q,m,n}|^2=|m_{\phi^c_Q,m,n}|^2
=\frac{(2\pi)^2}{\cV\,U_2}\,\,\Big|(n+\omega_2)-U(m+\omega_1)\Big|^2,
 \ \ m,n\in {\bf Z},
\label{qspectrum}
\eea
For the fermionic modes, the spectrum is 
\bea\label{fspectrum}
\qquad
m_{\psi_Q,m,n}^2=\,
\frac{(2\pi)^2}{\cV\,U_2}\,\,\big|n-U\,m\big|^2,
 \qquad  m,n\in {\bf Z},
\eea
since they do not acquire any  SU(2)$_R$ Scherk-Schwarz twist.

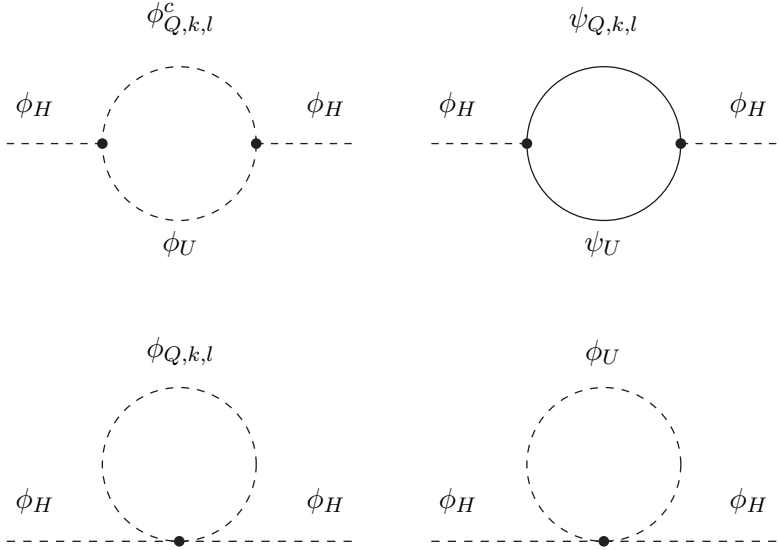
\begin{figure}[tbh]
\begin{center}
\begin{picture}(450,200)(-265,-150)
  \DashLine(-185,10)(-149,10){3} \Vertex(-149,10){2}
  \Text(-175,20)[b]{$\phi_H$}
  \DashCArc(-120,10)(29,0,360){3}
  \Text(-120,51)[b]{$\phi^c_{Q,k,l}$}
  \Text(-120,-23)[t]{$\phi_U$}
  \DashLine(-91,10)(-55,10){3} \Vertex(-91,10){2}
  \Text(-65,20)[b]{$\phi_H$}
  \DashLine(-25,10)(11,10){3} \Vertex(11,10){2}
  \Text(-15,20)[b]{$\phi_H$}
  \CArc(40,10)(29,0,360)
  \Text(40,51)[b]{$\psi_{Q,k,l}$}
  \Text(40,-23)[t]{$\psi_U$}
  \DashLine(69,10)(105,10){3} \Vertex(69,10){2}
  \Text(95,20)[b]{$\phi_H$}
  \DashLine(-185,-140)(-120,-140){3} \Text(-175,-130)[b]{$\phi_H$}
  \DashCArc(-120,-111)(29,0,360){3} \Vertex(-120,-140){2}
  \Text(-120,-74)[b]{$\phi_{Q,k,l}$}
  \DashLine(-120,-140)(-55,-140){3} \Text(-65,-130)[b]{$\phi_H$}
  \DashLine(-25,-140)(40,-140){3} \Text(-15,-130)[b]{$\phi_H$}
  \DashCArc(40,-111)(29,0,360){3} \Vertex(40,-140){2}
  \Text(40,-74)[b]{$\phi_U$}
  \DashLine(40,-140)(105,-140){3} \Text(95,-130)[b]{$\phi_H$}
\end{picture}
\caption{One-loop Feynman diagrams contributing to the Higgs mass.}
\label{self}
\end{center}
\end{figure}

With these considerations we can evaluate the one-loop correction to
$m_{\phi_H}(q^2)$ for non-zero external momentum $q^2$. 
As mentioned, if $H_u$ is a bulk field the correction below refers instead 
to the mass of the  zero mode of $\phi_H$.
In the component formalism, 
there are four one-loop Feynman diagrams contributing to the Higgs
mass as shown in Fig.~1. 
In the Euclidean space, the bosonic and fermionic contributions 
are respectively

\bea
\label{mass1}
-im^2_{\phi_H}(q^2)\Big|_{\cal B}&=&\!\!-i\,f_{4,t}^2N_c\,
\mu^{4-d}\sum_{m,n\in {\bf Z}}
\int \frac{d^dp}{(2\pi)^d}
\frac{(p+q)^2+p^2}{((p+q)^2+|m_{\phi_Q,m,n}|^2)\,\,  p^2}, 
\nonumber\\[9pt]
-im^2_{\phi_H}(q^2)\Big|_{\cal F}&=&i\, f_{4,t}^2N_c\, 
\mu^{4-d}\sum_{m,n\in {\bf Z}}
\int \frac{d^dp}{(2\pi)^d}
\frac{2p.(p+q)}{((p+q)^2+|m_{\psi_Q,m,n}|^2)\,\,p^2}.
\eea
where $\mu$ is the (finite, non-zero) mass scale introduced by the DR
scheme that we use  ($d\!=\!4\!-\!\epsilon$, $\epsilon\!\ra\! 0$).
If $q^2\!=\!0$,  mathematical consistency of eq.(\ref{mass1}) 
requires the introduction of an infrared (IR) mass shift  $m$
of the Kaluza-Klein masses in the fermionic denominator, to
ensure that the fermionic  contribution of the {\it massless} mode (0,0) 
does not vanish in the DR scheme used here\footnote{This vanishing
  would then be  an artifact of our DR regularisation.}. 
Supersymmetry then requires such IR mass shift be introduced 
for both  fermionic and bosonic contributions,
$|m_{\phi_{Q,m,n}}|^2\rightarrow |m_{\phi_{Q,m,n}}|^2+m^2$
and $|m_{\psi_{Q,m,n}}|^2\rightarrow|m_{\psi_{Q,m,n}}|^2+m^2$. 
Then,  a quartic divergence $m^4/\epsilon$ 
can be shown to be present in bosonic and fermionic contributions
to  $m^2_{\phi_H}(0)$, which cancels in their sum 
by initial supersymmetry\footnote{We
  return to this point later in the text.}. 
These arguments will become clear shortly from our more general
 computation of  $m^2_{\phi_H}(q^2)$, presented below.

After computing the momentum integrals in the DR scheme, 
one uses   the spectrum in eqs.(\ref{qspectrum}), 
(\ref{fspectrum}) to perform the summations to find the 
 one-loop result (with
 $\kappa_\epsilon\equiv (\mu \sqrt{\cV})^\epsilon$)

\bea
\label{mass2}
m^2_{\phi_H}(q^2)\Big|_{\cal B}\!\!\!
&=&\,
\frac{f_{4,t}^2\,N_c \,\kappa_\epsilon}{2\,\cV}
\!\int^1_0 \!
dx\,\bigg\{\frac{2-\epsilon/2}{\pi}\,{\cal J}_2[\omega_1,\omega_2,\delta]
+ \frac{q^2 \,\cV}{(2\pi)^2} \Big(x^2-x+\frac{1}{2}\Big)\,
{\cal J}_1[\omega_1,\omega_2,\delta]\bigg\},
\nonumber\\[12pt]
m^2_{\phi_H}(q^2)\Big|_{\cal F}\!\!
&=&\!\!-\frac{f_{4,t}^2\,N_c\,\kappa_\epsilon}{2 \,\cV}
\int^1_0 dx\,\bigg\{\frac{2-\epsilon/2}{\pi}\, {\cal J}_2[0,0,\delta]
+\frac{q^2\,\cV }{(2\pi)^2} \,x(x-1)\, {\cal J}_1[0,0,\delta]\bigg\}\\
\nonumber
\eea
The functions ${\cal J}_p$, $(p=1,2)$  have the following 
definition and leading behaviour in $\epsilon$

\bea
{\cal J}_p[\rho_1,\rho_2,\delta]&\equiv & \sum_{n_1,n_2\in {\bf Z}}
\int^\infty_0 \frac{dt}{t^{p-\epsilon/2}}\,\,
e^{-\pi t\,\big[\,\delta\,+\,\tau\, |(n_2+\rho_2)-U(n_1+\rho_1)|^2\,\big]}
\nonumber\\[10pt]
&=&\frac{(-\pi\, \delta)^p}{p \,\tau\, U_2}\bigg[\frac{2}{\epsilon}\bigg]
+\cJ_p^f[\rho_1,\rho_2,\delta]+ {\cal O}(\epsilon),\qquad\quad
\rho_{1,2}\in\bR;\,\,  p=1,2.
\label{jfunc1}\\
\nonumber
\eea
\noindent
where $\cJ_p^f$, $(p=1,2)$ 
are finite (no poles in $\epsilon$) and their  expression 
can be found in  Appendix~A, eqs.(\ref{a2}) to (\ref{a5}).
Note that the pole structure of $\cJ_{1,2}$  is {\it independent} of
$\rho_{1,2}$, which mark the differences between 
bosonic and fermionic spectra eqs.(\ref{qspectrum}), (\ref{fspectrum}).
  Further, in eq.(\ref{mass2})
\bea\label{deltatau}
\tau \equiv \frac{x}{U_2}, \qquad\quad
  \delta\equiv x\, (1-x)\,\, \frac{q^2\,\cV}{(2\pi)^2}
\eea
The  terms in eq.(\ref{mass2}) 
multiplied by a $q^2$ coefficient are absent
when considering $m^2_{\phi_H}(q^2=0)$, as can be
seen directly from eq.(\ref{mass1}) at $q^2=0$. 
In eq.(\ref{mass2}) the role of the aforementioned IR mass shift ($m^2$)
is played by the non-zero $q^2$.
The total one-loop mass correction which is the sum of the two
equations  in (\ref{mass2}) is then

\bea
m^2_{\phi_H}(q^2)\!\!\!
&=&\!\!\!
\frac{f_{4,t}^2 N_c}{\pi\,\cV}
\int^1_0 dx\,\bigg[{\cal J}_2[\omega_1,\omega_2, \delta]-{\cal
    J}_2[0,0,\delta ]\bigg]
\nonumber \\[10pt]
&+&\!
\frac{f_{4,t}^2 N_c \,\kappa_\epsilon}{8\pi^2\,}\,\, q^2
\!\int^1_0 \!\! 
 dx \bigg[  \Big(x\,(x-1)+\frac{1}{2}\Big)\,
{\cal J}_1[\omega_1,\omega_2,\delta] - x(x-1){\cal J}_1[0,0,\delta]\bigg].\qquad
\label{massyukawa}\\
\nonumber
\eea
with $\delta, \tau$ as in (\ref{deltatau}).
The second  line above is not present when 
computing $m_{\phi_H}(q^2\!=\!0)$.  
Using  eq.(\ref{jfunc1}), we find that the divergence
of ${\cal J}_2$ which is proportional to $\delta^2/\epsilon\sim
q^4/\epsilon$ is cancelled\footnote{This   divergence is nothing but the
$\frac{m^4}{\epsilon}$ discussed  after eq(\ref{mass1}),
cancelled in $m^2_{\phi_H}\!(0)$ by initial supersymmetry.}
between the bosonic and fermionic contributions
due to  equal number of bosonic and fermionic degrees of freedom 
(ensured by the initial supersymmetry).
Therefore, $m^2_{\phi_H}(0)$ given by the first line in 
(\ref{massyukawa}), is finite at the one-loop level.
There remains however a divergence in (\ref{massyukawa}) from the
${\cal J}_1$ functions, which although  have (according to
(\ref{jfunc1})) identical pole structure (UV behaviour), they have
a different coefficient ($x$ dependence) in front.
Then eq.(\ref{massyukawa}) gives
\bea\label{mass4}
m^2_{\phi_H}(q^2)=m^2_{\phi_H}(0)-\frac{f^2_{4,t}N_c}{16\pi}\, \,
\frac{q^4\cV}{(2\pi)^2}\,\,
\bigg[\frac{1}{\epsilon}+\ln\,(\mu \,\sqrt{\cV})\bigg]
+\frac{1}{\cV}\,{\cal O}(q^2\,\cV).
\eea
where ${\cal O}(q^2\,\cV)$ are finite terms originating from the 
(finite part of) $J_1$ functions and which can be evaluated
numerically using the formulae in the Appendix. Also, in (\ref{mass4})
\begin{eqnarray}\label{yukawa0}
m^2_{\phi_H}(0) &=&
\frac{f_{4,t}^2\,N_c}{2 \pi \cV}
\bigg\{-\frac{2}{3} \pi^2 U_2^2 \,
\Delta_{\omega_1}^2 \,(1- \Delta_{\omega_1})^2
\nonumber\\
\nonumber\\
&+&\!
\sum_{n_1\in\bZ}\bigg[\vert n_1+\omega_1\vert\, 
\Li_2 ( e^{-2 \pi\sigma_{n_1}})+\frac{1}{2\pi \,U_2}\,\,
\Li_3 ( e^{-2  \pi\sigma_{n_1}})\! +\!c.c.\!-\!(\omega_{1,2}\!\rightarrow\!
0)\bigg]\bigg\}
\qquad
\end{eqnarray}
where  $\Delta_{\omega_i}$ is the 
fractional part of
$\omega_i$, defined as $\Delta_{\omega_i}=\omega_i-[\omega_i]$, 
$0\leq \Delta_{\omega_i}<1$ and $[\omega_i]\in \bZ$, $i=1,2$.
Finally $\sigma_{n_1}=i\, 
[ \omega_2-U_1 (\omega_1+n_1)]+\, U_2 \,\vert n_1+\omega_1\vert$.

To conclude, in eq.(\ref{mass4})  a  quartic divergence
$q^4/\epsilon$  remains present at one-loop  
in the two point Green function  of the scalar field self-energy.
This  is of  {\it identical} type to that discussed above 
and cancelled by initial  supersymmetry in $m^2_{\phi_H}(0)$ 
(and also referred to as $m^4/\epsilon$).
In conclusion (initial) supersymmetry does not protect against the presence
of this (remaining) divergence and this finding questions
the power of (initial) supersymmetry in maintaining, after compactification,
a mild  UV behaviour of the scalar field mass. 

The presence of this divergence 
signals the need for a higher dimensional
(derivative) counterterm. This finding is not too surprising
if we recall that initial (6D)  theory, although supersymmetric, is
nevertheless non-renormalisable, where the presence of such operators
as loop counterterms can be expected. These findings extend
previous studies  for the case of 
$S_1/(Z_2\!\times\! Z_2')$ and $S_1/Z_2$ orbifolds
\cite{Ghilencea:2004sq,Ghilencea:2005hm,dghm} where similar
counterterms were found, regardless of the supersymmetry breaking mechanism
considered there. The counterterm which cancels the 
 pole in (\ref{mass4}) and respects the symmetries of the model is
 (if $H_u$ is a brane field)
\begin{eqnarray}\label{cterm1}
\int \! d^4 x \,d^2 \theta \, d^2\overline \theta
\,\lambda_t^2 \, H_u^\dagger \Box  H_u
\sim \! f_t^2 \!
\int \!\! d^4 x \,
\,\cV \, \,\phi_{H}^\dagger  \Box^2  \phi_{H}
+...
\end{eqnarray}
If $H_u$ is instead a bulk field, a similar result is obtained, and the
counterterm is
\begin{eqnarray}\label{cterm2}
\int \! d^4 x \,dy_1 \,dy_2 \!\int \! d^2 \theta \, d^2\overline \theta
\,\delta(y_1)\delta(y_2)\, \lambda_t^2 \, H_u^\dagger \Box  H_u
&\sim& \! f_t^2\!
\int\! d^4 x \,
\cV \!\!\! \sum_{k,l,m,n\in {\bf Z}} \phi_{H,k,l}^\dagger \Box^2  \phi_{H,m,n}
\nonumber\\
\nonumber\\
&\sim& \! f_t^2\!
\int \!\! d^4 x \,
\cV \, \phi_{H,0,0}^\dagger  \Box^2  \phi_{H,0,0}
+\cdots
\end{eqnarray}
The localisation of the counterterm is  justified by a simple but
powerful argument that if the counterterm were not localised, it should
have N=2 supersymmetry (rather than N=1)
 and thus would  necessarily involve the $H^c_u$ field. 
However $H^c_u$  has no Yukawa interaction, see
eq.(\ref{interaction}) and thus there is no bulk counterterm for 
(\ref{mass4}).

\subsection{One-loop gauge correction to a brane scalar field.}
\label{gauge-correction}

Now let us consider the  gauge correction to the mass of a brane 
scalar $\phi_H$.  Using eq.(\ref{gauge_int})
 the one-loop gauge correction  at zero external momentum,
in the DR scheme is (see also \cite{Choi:2003bh})
\bea
m^2_{\phi_H}(0)=-4\,g^2_4\, \mu^{4-d}\bigg\{
\!\sum_{n_1,n_2\in {\bf Z}}\int \frac{d^d p}{(2\pi)^d}
\frac{1}{p^2+|M_{n_1,n_2}(\omega_1,\omega_2)|^2} 
- \Big( \omega_{1,2}\ra 0\Big)
\bigg\}
\label{mass_gauge}
\eea
where $g_4=g_6/\sqrt{\cV}$, \,$d\!=\!4\!-\!\epsilon$ and the Kaluza-Klein 
mass spectrum is
\bea
|M_{n_1,n_2}(\omega_1,\omega_2)|^2&=&\frac{(2\pi)^2}{\cV\,U_2}\
 \vert(n_2+\omega_2)-U(n_1+\omega_1)\vert^2.
\eea
In eq.(\ref{mass_gauge}) we must introduce a small (but otherwise
arbitrary) mass shift $m^2$ of the denominators 
 to avoid ambiguities specific to the DR 
scheme that we are using. Indeed, $M_{0,0}(0,0)=0$ and in that
case the  integral in (\ref{mass_gauge}) with $\omega_i=0$
  would vanish for the $(0,0)$
Kaluza-Klein mode. A non-zero mass shift\footnote{This is similar to
  Yukawa corrections to $m_{\phi_H}^2(0)$, see also the text before 
eq.(\ref{mass2}), and eq.(\ref{yukawa0}).}
\begin{eqnarray}
|M_{n_1,n_2}(\omega_1,\omega_2)|^2\rightarrow 
|M_{n_1,n_2}(\omega_1,\omega_2)|^2+m^2
\end{eqnarray}
ensures the mathematical 
consistency of  eq.(\ref{mass_gauge}) and the steps taken in its
evaluation below. By supersymmetry, one must introduce this mass
shift in both denominators in eq.(\ref{mass_gauge}).
At the end of the calculation one  takes $m^2 \cV\rightarrow 0$. 
 After performing the integrals one obtains

\bea\label{eqmassb}
m^2_{\phi_H}(0)&=&
\frac{4 \,g^2_4 \,\kappa_\epsilon}{4\,\pi \,\cV}
\bigg\{{\cal J}_2\Big[0,0,\frac{m^2\cV}{(2\pi)^2}\Big]-
{\cal  J}_2\Big[\omega_1,\omega_2, \frac{m^2
    \cV}{(2\pi)^2}\Big]\bigg\}_{m^2 \,\cV\ra 0}
\nonumber\\
\nonumber\\
&=& \frac{4\, g_4^2}{4 \,\pi \cV}\,
\bigg\{\cJ^f_2\Big[0,0,\frac{m^2\cV}{(2\pi)^2}\Big]
+\frac{m^4 \cV^2}{(4\pi)^2\,\epsilon}- 
\cJ^f_2 \Big[\omega_1,\omega_2, \frac{m^2\cV}{(2\pi)^2}\Big]
-\frac{ m^4 \cV^2}{(4\pi)^2\,\epsilon}\bigg\}_{m^2 \,\cV\ra 0}
\!\!\!\!\!+\cO(\epsilon)
\nonumber\\
\nonumber\\
&=&\frac{4\,g_4^2}{4\, \pi \,\cV}\,\,
\bigg\{
\frac{2}{3}\,\pi^2\, U_2^2 \, \Delta_{\omega_1}^2 \,(1- \Delta_{\omega_1})^2
\nonumber\\
\nonumber\\
&-&
\sum_{n_1\in\bZ}\bigg[\vert n_1+\omega_1\vert\, 
\Li_2 ( e^{-2 \pi\sigma_{n_1}})+\frac{1}{2\pi \,U_2}\,\,
\Li_3 ( e^{-2  \pi\sigma_{n_1}})\! +\!c.c.\!-\!(\omega_{1,2}\!\rightarrow\!
0)\bigg]\bigg\}
\eea
where in the last step a notation identical to that after
eq.(\ref{yukawa0})
was used.
To obtain this result 
we used the definitions of $\cJ_{2}$ and $\cJ_{2}^f$
functions eq.(\ref{jfunc1})  with $\tau=1/U_2$ and $\delta=m^2
\cV/(2\pi)^2$.
It turns out that the  divergent part $m^4 \,\cV/\epsilon$
cancels out (due to initial supersymmetry)
 in the difference between bosonic and fermionic
contributions, and one finds a finite (i.e. no poles in $\epsilon$) 
one-loop correction to the mass of the brane scalar. The result in
(\ref{eqmassb}) agrees with that  in ref.~\cite{hmlee} for $U_1=0$.

To evaluate the momentum dependence of the
gauge correction to the mass of the brane scalar field, $m^2_{\phi_H}(q^2)$,
we shall use in the following the superfield 
formalism\footnote{One could also consider the momentum dependence of  gauge
 correction in component formalism as in \cite{hmlee}. However, the 
supersymmetric gauge fixing term would change the component field 
calculation. For this reason  we find it easier to work in the 
superspace formalism for the momentum-dependent part.}. 
For this purpose we compute
the gauge correction to the propagator of a (massless) brane
chiral multiplet in the absence of supersymmetry breaking.
To do so we need to consider only one supergraph with brane-chiral
and bulk-vector multiplets ``running'' in the loop.
We assume that, as in the 4D case, the soft (Scherk-Schwarz)
breaking  does not renormalize
the propagator of a massless brane chiral multiplet.
With an appropriate gauge fixing term, i.e. 
the 6D version of the super Feynman gauge,
the action for the vector superfield is
\be
S_6=\int d^4 x \,dy_1\, dy_2 \,d^2\theta\, d^2{\bar\theta}\,V[-\Box-\partial^2_5
-\partial^2_6] V.
\ee
Using this action, we compute the one-loop gauge correction
to the propagator of a brane chiral superfield (of 4-momentum $q$)
and located at the origin ($y_{1,2}=0$), which equals

\be
-\frac{g^2_6}{\cV}\int \frac{d^4 q}{(2\pi)^4}\,A(q)\,\bigg[\int d^4
\theta\,\,{\overline H}(-q,\theta)\,H(q,\theta)\bigg]
\ee
where
\bea
A(q)&=&\mu^{4-d}\sum_{n_1,n_2\in {\bf Z}}
\int\frac{d^dk}{(2\pi)^d}\frac{1}{(q+k)^2(k^2+M^2_{n_1,n_2}(0,0))}
\nonumber \\[9pt]
&=&\frac{\kappa_\epsilon}{(4\pi)^2} \int^1_0 \,dx
\sum_{n_1,n_2\in {\bf Z}}\int^\infty_0\frac{dt}{t^{1-\epsilon/2}}
\,\,e^{-\pi t(\delta+\tau|n_2-Un_1|^2)}
\eea
with $\epsilon=4-d$ and
\be\label{taudelta}
\kappa_\epsilon=(\mu\sqrt \cV)^\epsilon,\qquad
\tau=\frac{x}{U_2},\qquad 
\delta=x(1-x) \,\frac{q^2 \cV}{(2\pi)^2}
\ee
We use  the expression of $\cJ_1$ of eq.(\ref{jfunc1}) (see also
Appendix~A),  with $\tau$, $\delta$ given
in  (\ref{taudelta}), to find

\bea\label{wq1}
q^2\,A(q)&=& -\frac{1}{16\pi}\frac{q^4 \,\cV}{(2\pi)^2} 
\bigg[\frac{1}{\epsilon}+\ln(\mu\sqrt\cV)\bigg]\nonumber \\[9pt]
&&+\frac{q^2}{(4\pi)^2}
\int_0^1 dx  \,\cJ^f_1[0,0,x(1-x) \,q^2 \cV/(2\pi)^2]+\cO(\epsilon).\\
\nonumber
\eea
We thus find that a divergence $q^4\,\cV/\epsilon$ is generated  in the one-loop
corrected $q^2\,A(q)$ which is the coefficient (in momentum space) 
of the kinetic term of the brane scalar field,  component of the superfield $H$. 
Therefore, a higher derivative counterterm for the brane chiral multiplet
is needed in the tree-level action, to cancel the one-loop divergent term.
Its structure is similar to that found in (\ref{cterm1}).

In conclusion, similarly to the brane-localised superpotential
interactions, gauge interactions also generate, already at 
one-loop, higher derivative counterterms to the mass of the
brane scalar field. Note that one could consider a {\it formal} 
limit $\theta\ra 0$, when the two dimensions collapse onto each other.
Then  $\cV=(2\pi)^2 R_1 R_2 \sin\theta$ vanishes, 
the quartic divergence $q^4\,\cV/\epsilon$ 
present in eqs.(\ref{mass4}), (\ref{wq1}),  is not present
and   no higher derivative operators are generated at
one-loop for one extra dimension (although they will be generated at
higher orders)\footnote{This is not in contradiction with findings
in  5D case
\cite{Ghilencea:2004sq,Ghilencea:2005hm,dghm} where such operators
were nevertheless generated, since there  Yukawa 
 interaction involved two bulk fields
(unlike here where it has  one bulk field only).}.
However,  in order to keep the  6D (dimensionful)  
Yukawa and gauge couplings non-zero in a  5D limit, 
one must  keep the volume non-zero (and finite), which 
is the limit of a ``squeezed'' torus. As a result the
 higher derivative counterterms do not decouple in the 5D limit.

\section{Further Remarks and Conclusions.}\label{section5}

In the context of general 6D compactifications  on the orbifold
 $\cT^2/Z_2$ and with Scherk-Schwarz mechanism for supersymmetry breaking
 we investigated the one-loop corrections to the mass of a scalar
field,  induced by (brane localised) superpotentials and by (bulk)
gauge  interactions. 
Our results show that both  interactions usually generate,
after transmission of supersymmetry breaking,
higher derivative operators as one-loop counterterms to the mass of the 
scalar field. This is an important finding, in particular for the
hierarchy problem, since in such models the 
scalar field can be regarded as a Higgs field candidate.

Although such operators are higher dimensional (and thus suppressed by
the UV cutoff of initial theory, or in a 4D language by the volume)
 raising this scale suppresses
them at the classical level only. Indeed, we showed that
such operators are nevertheless generated  dynamically, already at
one-loop,  while integrating  out the bulk modes.
As a result of this, we find that the two-point Green
function  for the scalar field self-energy has, above the
compactification scale and  under the UV scaling of the 
external momentum $q\ra \lambda_s\, q$, a quartic dependence
on the scaling parameter $\lambda_s$. 
This happens despite the initial supersymmetry
of the higher dimensional theory and its soft (Scherk-Schwarz)
breaking and raises intriguing questions one the role of  initial
supersymmetry in enforcing a mild UV ``running''  
of scalar fields masses (above $1/R_{1,2}$).
Our technical results can also be used to investigate the running of
the scalar field mass  {\it across} the compactification scale,
from $q^2\ll 1/R_{1,2}^2$ to  $q^2\gg 1/R_{1,2}^2$.

The presence of such higher derivative counterterms to the scalar
fields masses is in the end not too surprising if we recall that
initial theory, although supersymmetric, is nevertheless
non-renormalisable and in these theories such operators
 are expected to be generated. At the technical level, the 
presence of these operators is related to the number of bulk fields in
the interaction. The origin of these operators 
is due, in our case, to a mixing effect between a {\it winding} zero-mode (on the 
lattice dual to that of Kaluza-Klein modes) wrt one dimension and 
the (infinite) series of Kaluza-Klein modes of the second dimension.
 This   indicates the {\it non-perturbative} and
non-local nature of the origin  of these counterterms. 
In the absence of an UV completion, the unknown coefficients (in the action) 
of such higher derivative counterterms  
prevent one from making predictions about physics at (momentum) 
scales at/above the compactification scales, despite the
supersymmetric nature of the uncompactified theory. 
Models  with higher derivative  operators may also 
raise potential conceptual problems, since in their presence 
further  complications  arise, such as unitarity violation, the
presence of additional ghost fields  or non-locality effects.

Our results extend to the case of two dimensional orbifolds, previous 
findings for  $S^1/Z_2$ and $S^1/(Z_2\times Z_2')$ compactifications
where brane-localised higher derivative operators are also
generated at  one-loop or beyond, regardless of the details of the 
supersymmetry breaking mechanism considered there. We expect that our results 
remain  valid for the case of other two-dimensional orbifolds
$\cT^2/Z_N$, $N>2$ and other mechanisms for (the 4D) supersymmetry breaking,
not considered here (like local breaking of supersymmetry).
Our argument to support this uses  that the origin of these operators is 
related to the number of bulk fields involved in the interaction and 
 to the details of the spectrum. These can be the main
differences from our $\cT^2/Z_2$. Assuming similar interactions
for $\cT^2/Z_N$ orbifolds,  the Kaluza-Klein  spectrum will only  
have different values for the parameters $\omega_{1,2}$ in the text, 
but as shown,  the pole structure of the functions $\cJ_{1,2}$, 
and thus the presence of higher derivative counterterms to the mass 
of the scalar field, is clearly independent of these parameters.

Higher derivative operators can have important implications for the scalar
potential and the cosmological constant in theories with compact
dimensions, although their exact role at the loop level 
was little investigated so far. Let us address some of the issues involved.
It is a common  approach in orbifold compactifications  to 
investigate the  one-loop corrections  to the scalar field masses and
their UV behaviour  by computing instead the
corresponding one-loop improved  scalar potential $\Lambda(\phi)$, using
$\Lambda(\phi)=\rm{Tr}\, \log\,\det(\Box+m^2(\phi))$. Here the trace
Tr is taken over all Kaluza-Klein modes associated with the
compactification.  A similar formula ``upgraded'' to respect string symmetries
(such as  modular invariance)  and to  include additional
string states \cite{Green}, is  also the
 starting point for the more comprehensive  string calculations.
 However, with this formula one can miss 
effects from higher derivative operators, which are {\it relevant}
at/above the
compactification scale(s).  Indeed,  the  second derivative of such
potential wrt scalar field $\phi$ only provides $m^2_\phi(0)$ and 
misses the effects of higher derivative operators which may be present
in the classical action (of an otherwise non-renormalisable theory) and
not accounted for in the above expression of $\Lambda(\phi)$.
To account for  this while still using the loop improved 
scalar potential approach, in the context of higher
dimensional theories, one  has to take account of higher derivative
operators already in the partition function of 
 the tree level action.  In this case the 
above expression for one-loop $\Lambda(\phi)$  is changed and is formally
given by $\Lambda(\phi)=\rm{Tr} \ln \det (\sigma \Box^2 \cV+
\Box+m^2(\phi))$  although the exact form 
can  further depend on other details of the
theory. Here $\sigma$ is the (unknown) coefficient of the higher derivative
operators in the tree level action.
In this expression  higher derivative operators provide the 
leading UV contribution, and thus must be included in the calculation 
of $\Lambda(\phi)$. The scalar field (physical) mass computed from
$\Lambda(\phi)$ then includes their effects too. These observations  also
 have implications for  the one-loop  cosmological constant in 
compactified theories.

We should mention that the higher derivative
operators on the visible brane can  ensure, when  included in the tree
level action, a better UV behaviour of 
the theory on the brane. This is because in their presence the
propagators change and, as a result  of this, loop corrections become 
less divergent or even finite (and this motivated in the past 
the use of such operators as regulators in 4D theories).  However the
loop corrections will then depend on the scale of such operators, 
which is the scale of ``new'' physics.  The presence of higher
derivative counterterms to scalar field masses after compactification 
of higher dimensional and {\it supersymmetric} theories, indicates
that,  in order  to solve the hierarchy problem,
one needs  a dynamical mechanism to fix  the 
scale of these  operators.

To conclude, we argued that
higher derivative operators are a common presence 
in compactifications of higher dimensional theories and are  radiatively
generated as counterterms to the scalar masses or the couplings in the theory. 
Their presence at the quantum 
level underlines the importance of their further investigation, with
 implications for  the hierarchy and cosmological 
constant problems. 

\vspace{0.3cm}
\noindent
{\bf Acknowledgements:}

\noindent
D.G. thanks Wilfried Buchm\"uller and the Theory Group at DESY (Hamburg) for the
warm hospitality and the  support which made possible his visit to DESY 
where part of  this work was done.
D.~Ghilencea is supported by a research grant from the Particle
Physics and Astronomy Research Council PPARC, U.K.

\section*{Appendix}

\def\theequation{\thesubsection-\arabic{equation}} 
\subsection*{A. Calculation of one-loop integrals on $\cT_2/Z_2$.}
\def\thesubsection{A} 
\setcounter{equation}{0}

In the text (eq.(\ref{jfunc1})) we used the series of regularised 
integrals  $\cJ_p$, for $p=1,2$ defined as 

\begin{eqnarray}\label{j1general}
\cJ_p[\rho_1,\rho_2,\delta] \,
  \equiv\! \sum_{n_1,n_2\in\bZ}
\int_0^\infty\! \!\!\frac{dt}{t^{p+\epsilon}}
\,\, e^{-\pi \, t\, \left[\,\delta+\, \tau\,
\vert n_2+\rho_2-U\,(n_1+\rho_1)\vert^2\,\right]},
\,\,\,\,\,  \tau, \delta >0;\, \rho_{1,2}\in\bR
\\
\nonumber
\end{eqnarray}
with $U\equiv U_1+i U_2$, $U_2>0$, $U_1\in\bR$. To recover 
eq.(\ref{jfunc1})  one must replace $\epsilon\ra -\epsilon/2$
in (\ref{j1general}).

This expression of $\cJ_p$, $p=1,2$  generalises previous
results in the Appendix of \cite{Ghilencea:2005hm}, 
 \cite{Ghilencea:2003xy}. The evaluation of $\cJ_{1,2}$ follows 
similar steps (see also Appendix B in \cite{Ghilencea:2005vm}).
 After a long calculation
one obtains that,  if $0\leq\delta/(\tau U_2^2)<1$

\begin{eqnarray}\label{a2}
\!\!\cJ_1[\rho_1,\rho_2,\delta] \!
& =&  \!\!\! \frac{\pi \delta}{\tau \,U_2} \bigg\{
\frac{1}{\epsilon}+ \ln\Big[4\pi \,(\tau U_2^2)\,                       
    e^{\gamma+\psi(\Delta_{\rho_1})+\psi(-\Delta_{\rho_1})}\Big] \bigg\}
\nonumber\\
\nonumber\\
&+& 2\pi U_2
\Big[\frac{1}{6}+\Delta_{\rho_1}^2-\Big(\delta/(\tau
    U_2^2)+\Delta_{\rho_1}^2\Big)^\frac{1}{2}\Big]
 -   \sum_{n_1\in\bZ}
\ln\Big\vert 1-e^{-2\pi \,\gamma (n_1)}\Big\vert^2
\nonumber\\
\nonumber\\
& +\!&\!\! \sqrt\pi\, U_2 
\!\sum_{p\geq 1}^\infty \frac{\Gamma[p\!+\!1/2]}{(p\!+\!1)!}
\bigg[\frac{-\delta}{\tau U_2^2}\bigg]^{p+1} 
\!\!
\Big(\zeta[2p\!+\!1,1\!+\! \Delta_{\rho_1}]\!+\!
\zeta[2p\!+\!1,1\!-\!\Delta_{\rho_1}]\Big)
\\
\nonumber
\end{eqnarray}
while if one has  $\delta/(\tau U_2^2)\geq 1$

\begin{eqnarray}\label{a3}
\cJ_1[\rho_1,\rho_2,\delta] 
&=&  \!\!\! \frac{\pi \delta}{\tau \,U_2} \bigg\{
\frac{1}{\epsilon}+\ln\Big[4\pi \,\delta \,
    e^{\gamma-1}\Big] \bigg\}
 -\!\sum_{n_1\in\bZ}\ln\Big\vert
 1\!-\!e^{-2\pi \,\gamma (n_1)}\Big\vert^2\,
\nonumber\\
\nonumber\\
\!&\!+&\!  4 \bigg[\frac{\delta}{\tau}\bigg]^{\frac{1}{2}}\!\!
\sum_{\tilde n_1>0}
\frac{1}{\tilde n_1}\,\, {\cos(2\pi \tilde n_1\,\rho_1)} \,\,
K_1\Big(2\pi\tilde n_1
(\delta/(\tau U_2^2))^{\frac{1}{2}}\Big)\qquad\qquad\qquad\qquad
\\
\nonumber
\end{eqnarray}
with the notation
\begin{eqnarray}\label{notation}
\gamma(n_1)&=&\frac{1}{\sqrt{\tau}} \,\big[z(n_1)\big]^{\frac{1}{2}}-i\,
(\rho_2-U_1(n_1+\rho_1))\nonumber\\
z(n_1)&=&\delta+\tau U_2^2(n_1+\rho_1)^2
\end{eqnarray}
Here          $\zeta[z,a]$ is the Hurwitz  
Zeta function, $\zeta[z,a]=\sum_{n\geq 0} (n+a)^{-z}$, $\rm{Re}\,
z>1$,\,$a\not=0,-1,-2,\cdots$, and  $\psi(x)=d/dx \,\ln \Gamma[x]$. 
Eqs.(\ref{a2}), (\ref{a3}) depend on fractional part of
$\rho_{1,2}\,$ defined by
$\Delta_{\rho_i} \equiv \rho_i-[\rho_i]$ with $0\!\leq\!
\Delta_{\rho_i}\!<\! 1$,
$[\rho_i]\in \bZ$. Finally, $K_n$ is the modified Bessel function \cite{gr} 
\begin{equation}\label{bessel1}
\int_{0}^{\infty} \! dx\, x^{\nu-1} e^{- b x^p- a
x^{-p}}=\frac{2}{p}\, \bigg[\frac{a}{b}
\bigg]^{\frac{\nu}{2 p}} K_{\frac{\nu}{p}}(2 \sqrt{a \, b}),\quad Re
(b),\, Re (a)>0
\end{equation}
with
\begin{eqnarray}
K_1[x]=e^{-x}\sqrt{\frac{\pi}{2
    x}}\left[1+\frac{3}{8 x}-\frac{15}{128 x^2}
+\cO(1/x^3)\right]
\\
\nonumber
\end{eqnarray}
which is  strongly suppressed at large argument.

One also finds that, if $\delta\ll \tau U_2^2$ and $\delta\ll 1$

\begin{eqnarray}\label{partj}
\cJ_1[\rho_1,\rho_2,\delta\ll 1] 
& =&  \frac{\pi \delta}{\tau \,U_2} \frac{1}{\epsilon}
-\ln\bigg\vert\frac{\vartheta_1(\rho_2-U \rho_1\vert U)}{(\rho_2-U
  \rho_1)\,
\eta(U) } \,e^{i
  \pi U \rho_1^2}\bigg\vert^2-\ln(\delta/\tau+\vert \rho_2-U \rho_1\vert^2)
\nonumber\\
\nonumber\\
\cJ_1[1/2,1/2,\delta\ll 1] 
& =&  \frac{\pi \delta}{\tau \,U_2} \frac{1}{\epsilon}
-\ln\bigg\vert\frac{\vartheta_1(1/2-U/2\vert U)}{
\eta(U) } \,e^{i
  \pi U/4}\bigg\vert^2
\nonumber\\
\nonumber\\
\cJ_1[0,0,\delta\ll 1] 
 &=&  \frac{\pi \delta}{\tau \,U_2} \frac{1}{\epsilon}
-\ln\Big[ \, 4\pi^2 \, \vert\eta(U) \vert^4 
\,\,\tau^{-1}\Big]-\ln\delta
\\
\nonumber
\end{eqnarray}
Above we used 
eq.(G-4) in \cite{Ghilencea:2003xy} and the notation

\begin{eqnarray}\label{ddt}
\eta(\tau) & \equiv & e^{\pi i \tau/12} \prod_{n\geq 1} (1- e^{2 i
\pi\tau\, n}),
\nonumber\\
\vartheta_1(z\vert\tau)&\equiv & 2 q^{1/8}\sin (\pi z) \prod_{n\geq 1} 
(1- q^n) (1-q^n e^{2 i \pi z}) (1- q^n e^{-2 i \pi z}), \qquad 
q\equiv e^{2 i \pi \tau}
\end{eqnarray}
In the following we also provide the result for
 $\cJ_2[\rho_1,\rho_2,\delta]$ whose calculation is 
similar.

\noindent
If $0\leq \delta/(\tau U_2^2)<1$
\begin{eqnarray}\label{part2j}
\!\!\cJ_2[\rho_1,\rho_2,\delta] \!
& =&  \!\!\! -\frac{\pi^2 \delta^2}{2\, \tau \,U_2} \frac{1}{\epsilon}
-\frac{\pi^2 \delta^2}{2 \,\tau\,U_2} \ln\Big[4\pi \,(\tau U_2^2)\,
    e^{\gamma+\psi(\Delta_{\rho_1})+\psi(-\Delta_{\rho_1})}\Big] 
\nonumber\\
\nonumber\\
&+& \pi^2 \tau U_2^3 \,\frac{1}{3}
\bigg\{
\frac{1}{15}-2\Delta_{\rho_1}^2 (1+ \Delta_{\rho_1}^2)
  -6\frac{\delta}{\tau U_2^2}\Big[\frac{1}{6}+\Delta_{\rho_1}^2\Big]
+4 \Big[\delta/(\tau
    U_2^2)+\Delta_{\rho_1}^2\Big]^\frac{3}{2}\bigg\}
\nonumber\\
\nonumber\\
&+&
\sum_{n_1\in\bZ}\left\{ \big(\tau \,z(n_1)\big)^{\frac{1}{2}}\,\,
\Li_2 ( e^{-2\pi\gamma(n_1)})+\frac{\tau}{2\pi}\,\,
\Li_3 ( e^{-2\pi\gamma(n_1)}) +c.c.\right\}
\nonumber\\
\nonumber\\
&+&
\!\! \pi^{3/2} \tau \, U_2^3 
\sum_{p\geq 1}^\infty \frac{\Gamma[p\!+\!1/2]}{(p\!+\!2)!}
\bigg[\frac{-\delta}{\tau U_2^2}\bigg]^{p+2} 
\!\!
\Big(\zeta[2p\!+\!1,1\!+\! \Delta_{\rho_1}]\!+\!
\zeta[2p\!+\!1,1\!-\!\Delta_{\rho_1}]\Big)
\\
\nonumber
\end{eqnarray}

\noindent
If instead $\delta/(\tau U_2^2)\geq 1$

\begin{eqnarray}\label{a5}
\!\!\cJ_2[\rho_1,\rho_2,\delta] \!
& =&  \!\!\! -\frac{\pi^2 \delta^2}{2 \,\tau \,U_2} \frac{1}{\epsilon}
-\frac{\pi^2 \delta^2}{2 \tau\,U_2} \ln\Big[\pi \,\delta\,
    e^{\gamma-3/2}\Big] 
\nonumber\\
\nonumber\\
&+&
\sum_{n_1\in\bZ}\left\{ \big(\tau \,z(n_1)\big)^{\frac{1}{2}}\,\,
\Li_2 ( e^{-2\pi\gamma(n_1)})+\frac{\tau}{2\pi}\,\,
\Li_3 ( e^{-2\pi\gamma(n_1)}) +c.c.\right\}
\nonumber\\
\nonumber\\
&+&
4\,\delta \,U_2 \sum_{\tilde n_1>0}\frac{1}{\tilde n_1^2}\,
\cos(2\pi\tilde n_1 \rho_1)\,K_2\Big(2\pi\,\tilde 
n_1(\delta/(\tau U_2^2))^{1/2}\Big)
\\
\nonumber
\end{eqnarray}
where we used the notation in eq.(\ref{notation}).

In the text we used the simpler case $\delta\ll 1$, 
$\delta/(\tau U_2^2)\ll 1$, when

\begin{eqnarray}
J_2[\rho_1,\rho_2,\delta\ll 1] &=&
-\frac{\pi^2 \delta^2}{2\, \tau \,U_2} \frac{1}{\epsilon}
+ \pi^2 \tau\, U_2^3 \,\frac{1}{3}\,
\Big[
\frac{1}{15}-2\Delta_{\rho_1}^2 (1- \Delta_{\rho_1})^2\Big]
\nonumber\\
\nonumber\\
&+&
\sum_{n_1\in\bZ}\left\{ \tau \,U_2\,\vert n_1+\rho_1\vert\, 
\Li_2 ( e^{-2 \pi\sigma_{n_1}})+\frac{\tau}{2\pi}\,\,
\Li_3 ( e^{-2  \pi\sigma_{n_1}}) +c.c.\right\}\qquad\quad
\end{eqnarray}
with $\sigma_{n_1}=i\, [ \rho_2-U_1 (\rho_1+n_1)]+\, U_2 \,
\vert n_1+\rho_1\vert$.
Also
\begin{eqnarray}
\!\!\!
J_2[0,0,\delta\ll 1] &=&
-\frac{\pi^2 \delta^2}{2\, \tau \,U_2} \frac{1}{\epsilon}
+\frac{\pi^2}{45}\,  \tau\, U_2^3 \,
\nonumber\\
\nonumber\\
&+&\!\!\! \frac{\tau}{2\pi} \!\!
\sum_{n_1\in\bZ}\Big\{2\pi \,U_2\,\vert n_1\vert\, 
\Li_2 ( e^{-2 \pi\sigma_{n_1}})+
\Li_3 ( e^{-2  \pi\sigma_{n_1}}) +c.c.\Big\}\qquad\qquad
\end{eqnarray} 
with $\sigma_{n_1}=U_2 \vert n_1\vert -i \,U_1 \,n_1$. Finally

\begin{eqnarray}
\!\!\!\!
J_2[1/2,1/2,\delta\ll 1] &=&
-\frac{\pi^2 \delta^2}{2\, \tau \,U_2} \frac{1}{\epsilon}
-\frac{7\pi^2}{360}\, \tau\, U_2^3 \,
\nonumber\\
\nonumber\\
&+&\!\!\!\frac{\tau}{2\pi} \!\!
\sum_{n_1\in\bZ}\Big\{ 2\pi \,U_2\,\vert n_1+1/2 \vert\, 
\Li_2 (- e^{-2 \pi\sigma_{n_1}})+
\Li_3 (- e^{-2  \pi\sigma_{n_1}}) +c.c.\Big\}\qquad
\\
\nonumber
\end{eqnarray}
with $\sigma_{n_1}=U_2 \vert n_1+1/2\vert- i U_1 (n+1/2)$.  
Setting $U_1\!=\!0$ in this section recovers the results 
of the Appendix in \cite{Ghilencea:2005hm}.



\begin{thebibliography}{99}


\def\apj#1#2#3{Astrophys.\ J.\ {\bf #1} (#2) #3}
\def\ijmp#1#2#3{Int.\ J.\ Mod.\ Phys.\ {\bf #1} (#2) #3}
\def\mpl#1#2#3{Mod.\ Phys.\ Lett.\ {\bf #1} (#2) #3}
\def\nat#1#2#3{Nature\ {\bf #1} (#2) #3}
\def\npb#1#2#3{Nucl.\ Phys.\ {\bf B #1} (#2) #3}
\def\plb#1#2#3{Phys.\ Lett.\ {\bf B #1} (#2) #3}
\def\prd#1#2#3{Phys.\ Rev.\ {\bf D #1} (#2) #3}
\def\prl#1#2#3{Phys.\ Rev.\ Lett.\ {\bf #1} (#2) #3}
\def\prt#1#2#3{Phys.\ Rep.\ {\bf #1}, (#2) #3}
\def\sjnp#1#2#3{Sov.\ J.\ Nucl.\ Phys.\ {\bf #1} (#2) #3}
\def\zp#1#2#3{Z.\ Phys.\ {\bf C #1} (#2) #3}
\def\jhep#1#2#3{JHEP \ {\bf #1} (#2) #3}
\def\epjc#1#2#3{Eur.\ Phys.\ J. \ {\bf C #1} (#2) #3}


\bibitem{antoniadis}
I.~Antoniadis,
  Phys.\ Lett.\ B {\bf 246} (1990) 377.

\bibitem{susyb}
E.~A.~Mirabelli and M.~E.~Peskin, 
{\it ``Transmission of supersymmetry breaking from a 4-dimensional boundary,''}
\prd{58}{1998}{065002}; 
  A.~Pomarol and M.~Quiros,
{\it  ``The standard model from extra dimensions,''}
  Phys.\ Lett.\ B {\bf 438} (1998) 255
  [arXiv:hep-ph/9806263]; 
I.~Antoniadis, S.~Dimopoulos, A.~Pomarol and M.~Quiros,
  {\it ``Soft masses in theories with supersymmetry breaking by
  TeV-compactification,''}
  Nucl.\ Phys.\ B {\bf 544} (1999) 503
  [arXiv:hep-ph/9810410]; 
K. A. Meissner, H. P. Nilles and M. Olechowski,
   {\it ``Brane induced supersymmetry breakdown and restoration,''}
   Acta Phys. Polon. {\bf B33} (2002) 2435; 
A.~Delgado, A.~Pomarol and M.~Quiros,
  {\it ``Supersymmetry and electroweak breaking from extra dimensions at the
  TeV-scale,''}
  Phys.\ Rev.\ D {\bf 60} (1999) 095008
  [arXiv:hep-ph/9812489]; 
K. A. Meissner, H. P. Nilles and M. Olechowski,
  {\it ``Supersymmetry breakdown at distant branes: The super-Higgs mechanism,''}
   \npb{561}{1999}{30}; 
D. E. Kaplan, G. D. Kribs and M. Schmaltz,
   {\it ``Supersymmetry breaking through transparent extra dimensions,''}
   \prd{62}{2000}{035010}; 
Z. Chacko, M. A. Luty, A. E. Nelson and E. Ponton,
   {\it ``Gaugino mediated supersymmetry breaking,''}
   \jhep{0001}{2000}{003}; 
J. A. Bagger, F. Feruglio and F. Zwirner, 
  {\it ``Generalized symmetry breaking on orbifolds,''}
   \prl{88}{2002}{101601};  
   {\it ``Brane induced supersymmetry breaking,''},
   \jhep{0202}{2002}{010}; 

\bibitem{equiv}
  D.~Marti and A.~Pomarol,
  {\it ``Supersymmetric theories with compact extra dimensions in N = 1
  superfields,''},  Phys.\ Rev.\ D {\bf 64} (2001) 105025
  [arXiv:hep-th/0106256]; 
D.~E.~Kaplan and N.~Weiner,
  {\it ``Radion mediated supersymmetry breaking as a Scherk-Schwarz
  theory,''} [arXiv: hep-ph/0108001]; 
A. Delgado, G. v. Gersdorff and M. Quiros,
{\it ``Brane-assisted Scherk-Schwarz supersymmetry breaking in orbifolds,''}
\jhep{0212}{2002}{002}.

\bibitem{nomura}
Y.~Nomura and D.~R.~Smith,
{\it ``Spectrum of TeV particles in warped supersymmetric grand unification,''}
 \prd{68}{2003}{075003}.

\bibitem{kl}
K.~Y.~Choi and H.~M.~Lee,
 {\it ``Softness of brane-localized supersymmetry breaking on
   orbifolds,''} \plb{575}{2003}{309}.

\bibitem{Delgado:2001si}
  A.~Delgado and M.~Quiros,
  {\it ``Supersymmetry and finite radiative electroweak breaking from an extra
  dimension,''}
  Nucl.\ Phys.\ B {\bf 607} (2001) 99
  [arXiv:hep-ph/0103058].
 
\bibitem{Derendinger:1985kk}
  J.~P.~Derendinger, L.~E.~Ibanez, and H.~P.~Nilles,
  {\it ``On The Low-Energy D = 4, N=1 Supergravity Theory Extracted From The D = 10,
  N=1 Superstring,''}
  Phys.\ Lett.\ B {\bf 155} (1985) 65.
  J.~P.~Derendinger, L.~E.~Ibanez, and H.~P.~Nilles,
  {\it ``On The Low-Energy Limit Of Superstring Theories,''}
  Nucl.\ Phys.\ B {\bf 267} (1986) 365.
  M.~Dine, R.~Rohm, N.~Seiberg, E.~Witten,
  {\it ``Gluino Condensation In Superstring Models,''}
  Phys.\ Lett.\ B {\bf 156} (1985) 55.

\bibitem{fll}
  A.~Falkowski, H.~M.~Lee and C.~Ludeling,
  {\it ``Gravity mediated supersymmetry breaking in six dimensions,''}
  arXiv:hep-th/0504091.


\bibitem{hmlee} H.~M.~Lee, 
{\it ``Softness of supersymmetry breaking on the orbifold
  $T^2$/Z(2),''},  \jhep{06}{2005}{044} [hep-th/0502093].

\bibitem{higgsmass}
N. Arkani-Hamed, L. Hall, Y. Nomura, D. Smith and N. Weiner,
{\it ``Finite radiative electroweak symmetry breaking from the bulk,''}
\npb{605}{2001}{81} [arXiv:hep-ph/0102090]; 
R.~Barbieri, L.~J.~Hall, Y.~Nomura,
{\it ``A constrained standard model from a compact extra dimension,''}
Phys.\ Rev.\ D {\bf 63} (2001) 105007
[arXiv:hep-ph/0011311]; 
R.~Barbieri, L.~J.~Hall, Y.~Nomura,
  {\it ``Softly broken supersymmetric desert from orbifold compactification,''}
  Phys.\ Rev.\ D {\bf 66} (2002) 045025
  [arXiv:hep-ph/0106190]; 
R.~Barbieri, L.~J.~Hall and Y.~Nomura,
  {\it``Models of Scherk-Schwarz symmetry breaking in 5D: Classification and
  calculability,''}
  Nucl.\ Phys.\ B {\bf 624} (2002) 63
  [arXiv:hep-th/0107004]; 
V.~Di Clemente, S.~F.~King and D.~A.~J.~Rayner,
  {\it ``Supersymmetry and electroweak breaking with large and small extra
  dimensions,''}
  Nucl.\ Phys.\ B {\bf 617} (2001) 71
  [arXiv:hep-ph/0107290]; 
V.~Di Clemente, S.~F.~King and D.~A.~J.~Rayner,
  {\it ``Supersymmetric Higgs bosons in a 5D orbifold model,''}
  Nucl.\ Phys.\ B {\bf 646}, 24 (2002)
  [arXiv:hep-ph/0205010]; 
  V.~Di Clemente and Y.~A.~Kubyshin,
  {\it ``Effective potential and KK-renormalization scheme in a 5D  supersymmetric
  theory,''}
  Nucl.\ Phys.\ B {\bf 636} (2002) 115
  [arXiv:hep-th/0108117];
  A.~Delgado, A.~Pomarol and M.~Quiros,
  {\it ``Supersymmetry and electroweak breaking from extra dimensions at the
  TeV-scale,''}
  Phys.\ Rev.\ D {\bf 60} (1999) 095008
  [arXiv:hep-ph/9812489].

\bibitem{Ghilencea:2001bw}
  D.~M.~Ghilencea, S.~Groot Nibbelink and H.~P.~Nilles,
  {\it ``Gauge corrections and FI-term in 5D KK theories,''}
  Nucl.\ Phys.\ B {\bf 619} (2001) 385
  [arXiv:hep-th/0108184].

\bibitem{Ghilencea:2004sq}
  D.~M.~Ghilencea,
  {\it ``Higher derivative
operators as loop counterterms in one-dimensional field
  theory orbifolds,''},  JHEP {\bf 0503} (2005) 009
    arXiv:hep-ph/0409214.

\bibitem{Ghilencea:2005hm}
  D.~M.~Ghilencea and H.~M.~Lee,
{\it ``Higher derivative operators from transmission of supersymmetry
  breaking on S(1)/Z(2),''}
  arXiv:hep-ph/0505187.

\bibitem{Oliver:2003cy}
J.~F.~Oliver, J.~Papavassiliou and A.~Santamaria,
{\it ``Can power corrections be reliably computed in models with extra
dimensions?,''}
Phys.\ Rev.\ D {\bf 67} (2003) 125004
[arXiv:hep-ph/0302083].
  D.~M.~Ghilencea,
  {\it ``Compact dimensions and their radiative mixing,''}
  Phys.\ Rev.\ D {\bf 70} (2004) 045018
  [hep-ph/0311264].
  S.~Groot Nibbelink and M.~Hillenbach,
  {\it ``Renormalization of supersymmetric gauge theories on
  orbifolds: Brane gauge
  couplings and higher derivative operators,''}
  Phys.\ Lett.\ B {\bf 616} (2005) 125
  [hep-th/0503153].

\bibitem{Pais} See for example:
  A.~Pais and G.~E.~Uhlenbeck,
{\it ``On Field Theories With Nonlocalized Action,''}
  Phys.\ Rev.\  {\bf 79} (1950) 145; 
S. W. Hawking, {\it ``Quantum Field theory and Quantum Statistics:
Essays in Honour of the 60 th Birthday of E.S. Fradkin''},
eds. A.~Batalin et al, Bristol, UK (1987); 
S.~W.~Hawking and T.~Hertog,
{\it ``Living with ghosts,''}
Phys.\ Rev.\ D {\bf 65} (2002) 103515
[hep-th/0107088]. 
  E.~Bergshoeff, M.~Rakowski and E.~Sezgin,
  {\it ``Higher Derivative Superyang-Mills Theories,''}
  Phys.\ Lett.\ B {\bf 185} (1987) 371; 
J.~Z.~Simon,
{\it ``Higher Derivative Lagrangians, Nonlocality, Problems And Solutions,''}
Phys.\ Rev.\ D {\bf 41} (1990) 3720; 
A.~V.~Smilga,
{\it ``Ghost-free higher-derivative theory,''}
hep-th/0503213; 
A.~V.~Smilga,
{\it ``Benign vs. malicious ghosts in higher-derivative theories,''}
Nucl.\ Phys.\ B {\bf 706} (2005) 598
[hep-th/0407231]; 
F.~J.~de Urries and J.~Julve,
{\it ``Ostrogradski formalism for higher-derivative scalar field theories,''}
J.\ Phys.\ A {\bf 31} (1998) 6949
[hep-th/9802115]; 
P.~D.~Mannheim and A.~Davidson,
{\it ``Dirac quantization of the Pais-Uhlenbeck fourth order oscillator,''}
hep-th/0408104; 
P.~D.~Mannheim and A.~Davidson,
{\it ``Fourth order theories without ghosts,''}
hep-th/0001115.

\bibitem{Arkani-Hamed:2001tb}
  N.~Arkani-Hamed, T.~Gregoire and J.~Wacker,
  {\it ``Higher dimensional supersymmetry in 4D superspace,''}
  JHEP {\bf 0203} (2002) 055
  [arXiv:hep-th/0101233].
  S.~J.~Gates, M.~T.~Grisaru, M.~Rocek and W.~Siegel,
  {\it ``Superspace, Or One Thousand And One Lessons In Supersymmetry,''}
  Front.\ Phys.\  {\bf 58} (1983) 1
  [arXiv:hep-th/0108200].

\bibitem{ss}
J.~Scherk and J.~H.~Schwarz,
  {\it ``Spontaneous Breaking Of Supersymmetry Through Dimensional Reduction,''}
  Phys.\ Lett.\ B {\bf 82} (1979) 60;
J.~Scherk and J.~H.~Schwarz,
  {\it ``How To Get Masses From Extra Dimensions,''}
  Nucl.\ Phys.\ B {\bf 153} (1979) 61.

\bibitem{leeanomaly}
H.~M.~Lee, work in progress.

\bibitem{Hamidi:1986vh}
  S.~Hamidi and C.~Vafa,
  {\it ``Interactions On Orbifolds,''}
  Nucl.\ Phys.\ B {\bf 279} (1987) 465; 
  L.~J.~Dixon, D.~Friedan, E.~J.~Martinec and S.~H.~Shenker,
  {\it ``The Conformal Field Theory Of Orbifolds,''}
  Nucl.\ Phys.\ B {\bf 282} (1987) 13.

\bibitem{Antoniadis:1993jp}
  I.~Antoniadis and K.~Benakli,
  {\it ``Limits on extra dimensions in orbifold compactifications of superstrings,''}
  Phys.\ Lett.\ B {\bf 326} (1994) 69
  [arXiv:hep-th/9310151]; 
  I.~Antoniadis and K.~Benakli,
  {\it ``Limits on the size of extra-dimensions,''}
  arXiv:hep-ph/0004240.
  I.~Antoniadis and K.~Benakli,
  {\it ``Large dimensions and string physics in future colliders,''}
  Int.\ J.\ Mod.\ Phys.\ A {\bf 15} (2000) 4237
  [arXiv:hep-ph/0007226].

\bibitem{kawamura}
  Y.~Kawamura,
  {\it ``Triplet-doublet splitting, proton stability and extra dimension,''}
  Prog.\ Theor.\ Phys.\  {\bf 105} (2001) 999
  [arXiv:hep-ph/0012125].

\bibitem{kkl}
  H.~D.~Kim, J.~E.~Kim and H.~M.~Lee,
 {\it ``Top-bottom mass hierarchy, s - mu puzzle and gauge coupling unification
 with split multiplets,''}
  Eur.\ Phys.\ J.\ C {\bf 24} (2002) 159
  [arXiv:hep-ph/0112094].

\bibitem{abc}
  T.~Asaka, W.~Buchmuller and L.~Covi,
 {\it ``Gauge unification in six dimensions,''}
  Phys.\ Lett.\ B {\bf 523} (2001) 199
  [arXiv:hep-ph/0108021].

\bibitem{gs}
  S.~Groot Nibbelink, H.~P.~Nilles, M.~Olechowski and M.~G.~A.~Walter,
  {\it ``Localized tadpoles of anomalous heterotic U(1)'s,''}
  Nucl.\ Phys.\ B {\bf 665} (2003) 236
  [arXiv:hep-th/0303101]; 
  G.~von Gersdorff and M.~Quiros,
  {\it ``Localized anomalies in orbifold gauge theories,''}
  Phys.\ Rev.\ D {\bf 68} (2003) 105002
  [arXiv:hep-th/0305024]; 
  H.~M.~Lee, H.~P.~Nilles and M.~Zucker,
  {\it ``Spontaneous localization of bulk fields: The six-dimensional case,''}
  Nucl.\ Phys.\ B {\bf 680} (2004) 177
  [arXiv:hep-th/0309195].

\bibitem{dghm}
D.~M.~Ghilencea, H.~M.~Lee, {\it ``Higher derivative operators from
  orbifold compactifications''}, to be published.

\bibitem{Choi:2003bh}
  K.~Y.~Choi, J.~E.~Kim and H.~M.~Lee,
  {\it ``Towards 5D grand unification without SUSY flavor problem,''}
  JHEP {\bf 0306} (2003) 040
  [arXiv:hep-ph/0303213].

\bibitem{Green}
M.~B.~Green, J.~H.~Schwarz, E.~Witten, {\it ``Superstring Theory''},
vol.2 (Chapter 8), Cambridge University Press, 1987.

\bibitem{Ghilencea:2003xy}
  D.~M.~Ghilencea,
  {\it ``Regularisation techniques for the radiative corrections of the
  Wilson lines and Kaluza-Klein states,''}
  Phys.\ Rev.\ D {\bf 70} (2004) 045011
  [arXiv:hep-th/0311187].

\bibitem{Ghilencea:2005vm}
  D.~M.~Ghilencea, D.~Hoover, C.~P.~Burgess and F.~Quevedo,
  {\it ``Casimir energies for 6D supergravities compactified on T(2)/Z(N) with
  Wilson lines,''}  arXiv:hep-th/0506164.

\bibitem{gr}
I.S. Gradshteyn, I.M.Ryzhik, {\it ``Table of Integrals, Series and
Products''}, Academic Press Inc., New York/London,  1965.

\end{thebibliography}
\end{document}